\documentclass[a4paper]{VisionStyle}

\usepackage{epsfig} 
 
\begin{document} 
 
\title{THE IMPRINT OF PRESUPERNOVA WINDS ON SUPERNOVA REMNANT EVOLUTION:  
TOWARDS MORE REALISTIC MODELS FOR TYPE Ia SUPERNOVA REMNANTS AND THEIR  
SPECTRA} 
 
\author{C.\,Badenes \and E.\,Bravo} 
 
\institute{ 
Dpt. Physics and Nuclear Engineering, Univ.Polit.Catalunya, Diagonal 647,  
08028 Barcelona, Spain 
\and  
Institute for Space Studies of Catalonia, Gran Capit\`a 2-4, 08034  
Barcelona, Spain } 
 
\maketitle  
 
\begin{abstract} 

Supernova remnants are usually analysed in the light of
hydrodynamical models of the interaction of supernova ejecta with either a
constant density ambient medium or a circumstellar medium produced by a constant
presupernova wind. However, the ejection of energetic wind during the
presupernova phase changes the ambient medium structure and, consequently, the
early supernova remnant evolution. We have analysed the evolution of young
remnants of type Ia supernovae, focusing on the imprint of the presupernova 
wind history on the supernova remnant structure and on the influence of the
explosion mechanism. We have found that the remnant evolution is most 
sensitive to
the explosion mechanism at ages not larger than a few hundred years, while the
presupernova history shows its influence at later epochs, before the Sedov
phase sets in.
 
 
\end{abstract} 
 
\section{Introduction} 
 
Supernovae have become one key piece in the knowledge building of modern
astronomy. They stand at the cornerstone of many fields, including cosmology,
galactic and stellar evolution, cosmic rays generation, and many others.
Type Ia supernovae (SNIa) are currently used as practical tools to measure
cosmological distances and obtain insight about what kind of Universe we are in.
However, the theoretical
knowledge of SNIa is still plagued by uncertainties about 
the nature of their progenitor systems and the nature of the 
explosion mechanism. 

Modern X-ray
observatories, like {\sl Chandra} and {\sl XMM-Newton} give the opportunity to
collect a huge amount of high quality, high resolution, data about supernova
remnants (SNRs). 
Whereas the atomic
codes that allow to compute and fit the X-ray spectra have experienced a
considerable improvement in the last years, this is not the case of our
knowledge of the hydrodynamical evolution of SNRs. There are two 
desirable directions of advance for our theoretical understanding of SNR
evolution: one is multi-dimensional hydrodynamical simulations, and the other is
one-dimensional hydrodynamical calculations, that allow more realistic and
more detailed supernova (SN) models. Here, we have focused on the last kind of studies.

Presupernova evolution is often neglected in the hydrodynamic simulations of 
type Ia SNR, because the mass loss episodes of the progenitors are not believed to  
have a deep influence on their circumstellar medium (CSM). 
The shaping of the CSM through stellar winds and the formation of wind  
bubbles have been studied extensively for massive early type stars (i.e.,  
type II SN progenitors), albeit only for constant mass loss episodes or  
combinations of them (\cite{ebravo-B2:wea77}, \cite{ebravo-B2:che89}). 
Up to now, the usual assumptions for the CSM surrounding type Ia SN  
progenitors have been either a constant density interstellar medium (ISM)  
or a power law CSM merging smoothly with the ISM. 
However, recent models of type Ia presupernova evolution predict  
substantial non-constant mass loss episodes during the binary evolution 
(\cite{ebravo-B2:hac96},\cite{ebravo-B2:lan00}). Evidence for this kind of 
behavior has been found in X-rays even for type II events (\cite{ebravo-B2:imm01}).
Light echoes have been detected around two overluminous type Ia supernovae: SN 1991T
(\cite{ebravo-B2:sch94}) and SN 1998bu (\cite{ebravo-B2:cap01}), which have been
attributed to the pressence of circumstellar dust.
Thus, it is now necessary to evaluate the influence of such non-constant 
presupernova winds on the properties of the SNR. 
 
The explosion mechanism of SNIa is still a matter of strong  
debate. 
Since the evolution of young SNRs is dominated by the SN ejecta profile,  
different explosion mechanisms (implying different ejecta profiles)
can lead to important variations in the remnant  
behavior at early stages.   
\cite*{ebravo-B2:dwa98} found  
that most ejecta profiles could be adjusted by an exponential profile,  
but also that this approach fails to reproduce many features of SNR  
evolution. Their results call for a more thorough study of the influence of the 
explosion mechanism on the properties of SNRs.
 
\section{Models and results}
 
\begin{table*}[!t] 
  \caption{
  Charateristics of wind models.
  } 
  \label{ebravo-B2_tab1} 
  \begin{center} 
    \leavevmode 
    \footnotesize 
    \begin{tabular}[h]{cccccccc} 
      \hline \\[-5pt] 
        & $v_{\rm w}$ & $M_{\rm w}$ & $K_{51}$ & $t_{\rm end}$ & $t_{\rm SN}$ &
	$R_{\rm s}$ & $R_{\rm s,Weaver}$ \\
      Wind model & (km s$^ {-1}$) & ($M_{\sun}$) & ($\times10^{51}$~ergs) &
      (Myr) & (Myr) & ($\times10^{19}$~cm) & ($\times10^{19}$~cm) \\[+5pt]
      \hline \\[-5pt] 
      A & 200 & 0.2 & $8\times10^{-5}$ & 0.2 & 0.7 & 3.2 & 2.6 \\
      B & 20 & 0.2 & $8\times10^{-7}$ & 0.2 & 0.7 & 1.4 & 1.0 \\
      C & 200 & 0.6 & $2.4\times10^{-4}$ & 1.5 & 1.5 & 5.0 & 4.5 \\
      D & 20 & 0.6 & $2.4\times10^{-6}$ & 1.5 & 1.5 & 2.0 & 1.8 \\
      \hline \\ 
      \end{tabular} 
  \end{center} 
  \begin{list}{}{}
	\item
  	NOTE.- $v_{\rm w}$ is the wind velocity, $M_{\rm w}$ is the total mass 
	ejected in the wind, and $K_{51}$ is the total kinetic energy of the 
	wind; $t_{\rm end}$ is the duration of the wind phase, and $t_{\rm SN}$
	is the time of the SN explosion; $R_{\rm s}$ and $R_{\rm s,Weaver}$ are
	the radius of the outer shell at the time of SN explosion, according to
	the present calculations and to \cite*{ebravo-B2:wea77},
	respectively.
  \end{list}
\end{table*} 

\begin{figure}[!bt] 
  \begin{center} 
    \epsfig{file=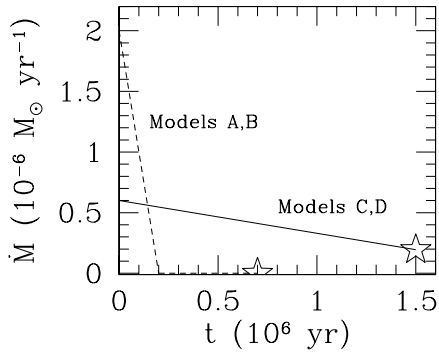, width=6cm} 
  \end{center} 
\caption{
Time evolution of wind mass-loss rate for models A and B ({\sl dashed line}) and
models C and D ({\sl solid line}). The time at which the SN explodes is
identified by a star. Note that in models A and B the wind ends 0.5~Myr before
SN explosion. 
}   
\label{ebravo-B2_fig1} 
\end{figure}

\begin{figure}[!bt] 
  \begin{center} 
    \epsfig{file=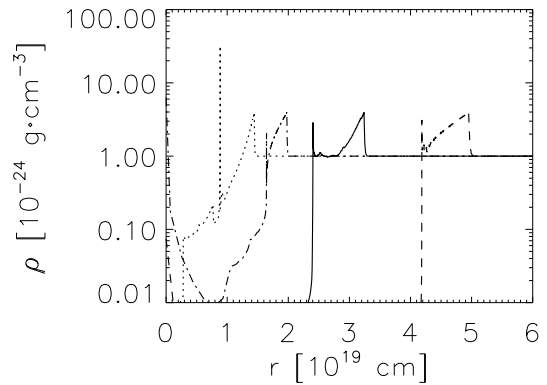, width=7cm} 
  \end{center} 
\caption{
Circumstellar medium density profile for each wind model at the time of SN
explosion. Lines correspond to wind models A ({\sl solid line}), B ({\sl dotted
line}), C ({\sl dashed line}), and D ({\sl dash-dotted line}).
}   
\label{ebravo-B2_fig2} 
\end{figure}

We have followed with an hydrodynamical code the interaction of the presupernova wind
with the surrounding ISM up to the time of SN explosion. 
The code was the same as in \cite*{ebravo-B2:bad01}, modified to take into
account the radiative losses, which are dynamically meaningful during the 
wind-ISM interaction. We have computed four
cases corresponding to different wind models, whose characteristics can be found in 
Table \ref{ebravo-B2_tab1} and Figure \ref{ebravo-B2_fig1} (the ISM was always assumed
to be initially homogeneous).
Figure \ref{ebravo-B2_fig2} displays the CSM  
profile for each model at the time of the SN explosion. 
 
\begin{figure}[!tb] 
  \begin{center} 
    \epsfig{file=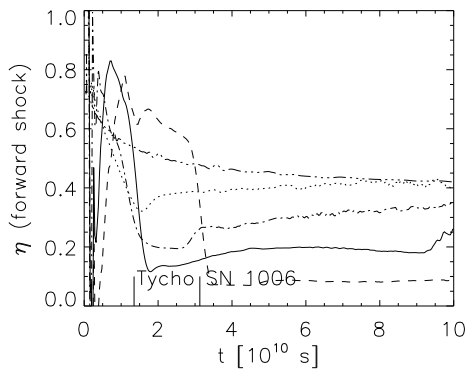, width=7cm} 
  \end{center} 
\caption{
Expansion parameter of the forward shock as a function of time for the
interaction of an exponential ejecta profile with the CSM as modifed by the four
wind models (see Fig.\ref{ebravo-B2_fig2}), together with the interaction of the
exponential ejecta with a constant density ISM ({\sl triple-dot-dashed
line}). The ages of Tycho's SNR and SNR 1006 are marked above the horizontal
axis.
}   
\label{ebravo-B2_fig3} 
\end{figure}

\begin{figure}[!tb] 
  \begin{center} 
    \epsfig{file=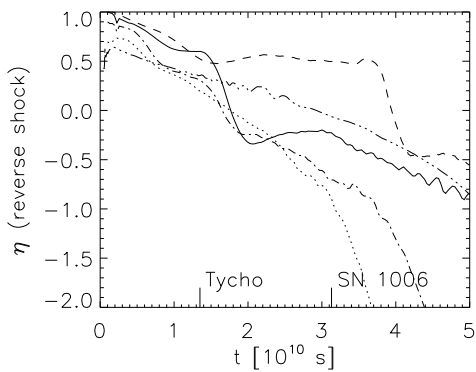, width=7cm} 
  \end{center} 
\caption{
The same as Fig.\ref{ebravo-B2_fig3} but for the reverse shock.
}   
\label{ebravo-B2_fig4} 
\end{figure}

The results of the hydrodynamic computation of wind-ISM interaction have been used as
the start configuration for the hydrodynamic calculation of SNR evolution. In this
study of the influence of different wind models we have used an analytic ejecta profile
given by an exponential law. The results are shown in Figures \ref{ebravo-B2_fig3} and 
\ref{ebravo-B2_fig4}, together with the  
evolution of the same ejecta profile interacting with a constant density  
ISM. We use the shock expansion parameters ($\eta \equiv {\rm d}\ln r/{\rm d}\ln t$)  
of the forward and reverse shocks as a figure of merit for model  
comparison. It can be seen that the
wind ejected by the progenitor system of a SNIa can have a  
dramatic influence on the dynamical evolution of its SNR. This evolution  
cannot be accurately described with the usual assumptions of a power law  
CSM merging smoothly with the ISM or direct interaction between the  
ejecta and the ISM (see also \cite{ebravo-B2:bad01}).

\begin{figure}[!bt] 
  \begin{center} 
    \epsfig{file=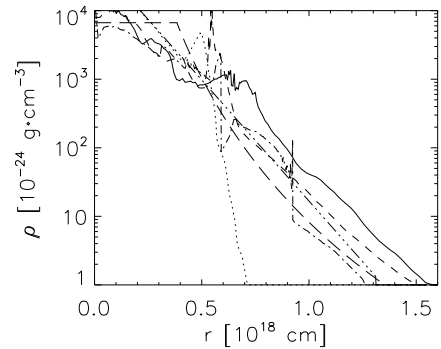, width=7cm} 
  \end{center} 
\caption{
Comparison of the ejecta density structure from different explosion models of
SNIa with the exponential ejecta profile ({\sl triple-dot-dashed line}), and a 
power-law profile with index 7 ({\sl long-dashed line}), at time $5\times10^8$~s 
after the explosion. Each of the represented SNIa models is based on a different 
explosion mechanism: deflagration ({\sl dotted line}), delayed detonation ({\sl solid
line}), pulsating delayed detonation ({\sl dashed line}), and
sub-Chandrasekhar ({\sl dash-dotted line}).
}   
\label{ebravo-B2_fig5} 
\end{figure}

We have also made a preliminary study of the influence of the explosion model
on the SNR properties. 
We have chosen a few representative models of SNIa explosion (see Table
\ref{ebravo-B2_tab2}): a  
delayed detonation (DDT), a pulsating delayed detonation (PDD), a  
deflagration (DEF) and a sub-Chandrasekhar explosion (SCH). In Figure 
\ref{ebravo-B2_fig5}, we compare  
them with the most usual analytical profiles: an exponential (EXP) and a  
power law with index 7 (PWL), both with ejected mass $1.4~M_{\sun}$ and 
kinetic energy $10^{51}$~erg.  
   
\begin{table*}[!t] 
  \caption{
  Characteristics of the SNIa explosion models.
  } 
  \label{ebravo-B2_tab2} 
  \begin{center} 
    \leavevmode 
    \footnotesize 
    \begin{tabular}[h]{ccccccccc} 
      \hline \\[-5pt] 
       Explosion & $M_{\rm ej}$ & $K_{51}$ & $t'$ &
       $M_{\rm Fe}$ & $M_{\rm Ca}$ & 
       $M_{\rm S}$ & $M_{\rm Si}$ & $M_{^{44}{\rm Ti}}$ \\
       model & ($M_{\sun}$) & ($\times10^{51}$~ergs) & ($10^9$~s) &
       ($M_{\sun}$) & 
       ($M_{\sun}$) & ($M_{\sun}$) & ($M_{\sun}$) & ($M_{\sun}$) \\[+5pt] 
      \hline \\[-5pt] 
      DEF & 1.38 & 0.79 & 11.4 & 0.68 & 0.006 & 0.022 & 0.031 & $4.7\times10^{-6}$ \\
      DDT & 1.37 & 2.00 & 7.1 & 1.04 & 0.023 & 0.069 & 0.084 & $5.6\times10^{-5}$ \\
      PDD & 1.36 & 1.52 & 8.2 & 0.83 & 0.020 & 0.070 & 0.090 & $1.1\times10^{-5}$ \\
      SCH & 0.97 & 0.98 & 7.7 & 0.49 & 0.027 & 0.102 & 0.141 & $1.2\times10^{-3}$ \\
      \hline \\ 
      \end{tabular} 
  \end{center} 
  \begin{list}{}{}
	\item
	NOTE.- $M_{\rm ej}$ and $K_{51}$ are, respectively, the ejected mass and
	the total kinetic energy released, and $t'$ is the normalization 
	timescale (see Figs. \ref{ebravo-B2_fig6} and \ref{ebravo-B2_fig7}). 
	The rest of columns give the yields
	of selected species. See \cite*{ebravo-B2:bra96} for more details on the 
	models.
  \end{list}
\end{table*} 
 
\begin{figure}[!tb] 
  \begin{center} 
    \epsfig{file=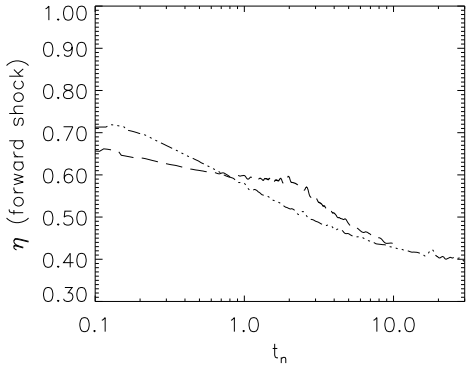, width=7cm} 
  \end{center} 
\caption{
Expansion parameter of the forward shock as a function of time for the
interaction of the analytical ejecta profiles with a constant
density ISM. Due to the different ejected masses and kinetic energies of the
models, the time scale has been normalized as in 
Dwarkadas \& Chevalier (1998). 
The normalization timescale is $10.4\times10^9$~s.
}   
\label{ebravo-B2_fig6} 
\end{figure}

\begin{figure}[!bt] 
  \begin{center} 
    \epsfig{file=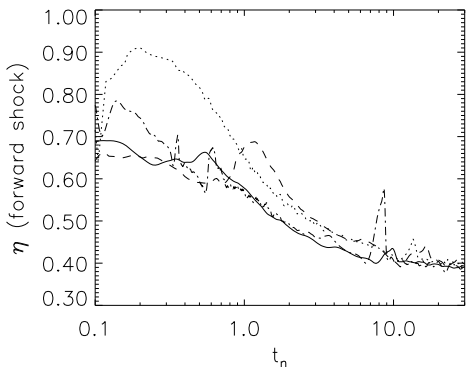, width=7cm} 
  \end{center} 
\caption{
The same as Fig. \ref{ebravo-B2_fig6} but for the numerical ejecta profiles (DEF,
DDT, PDD, and SCH). Table
\ref{ebravo-B2_tab2} gives the corresponding normalization timescales.
}   
\label{ebravo-B2_fig7} 
\end{figure}

The interaction of the different explosion models with a homogeneous ISM has been 
followed with the same hydrodynamical code. 
The evolution of the forward shock expansion parameter is given in Figures 
\ref{ebravo-B2_fig6} and \ref{ebravo-B2_fig7} (the  
evolution of the reverse shock expansion parameter shows only minor  
deviations between the different models and is not shown). 
It can be seen that the dynamical behavior of SNRs resulting from realistic SNIa 
explosion  
models cannot be accurately described either by a power law or an  
exponential density profile for the ejecta. Instead, the simulated dynamics show a  
mixture of features from both these simple analytic ejecta profiles 
(see also \cite{ebravo-B2:bad02}).

\section{Discussion and conclusions}  

We have shown that the CSM structure produced by different presupernova wind 
models can be discriminated 
against an initially (at SN explosion) uniform ISM through  
the dynamical properties of the SNR. The expansion parameter is most sensitive 
to the presupernova history at SNR ages larger than that of Tycho. 
We have also shown that different explosion models for SNIa can be discriminated 
through the  
dynamical properties of the  
SNR, and through the different chemical abundances of freshly synthesized  
elements (Fe, Ca and intermediate mass elements for X-ray lines, 44Ti for  
$\gamma$-ray lines). The expansion parameter is most sensitive to  
the explosion mechanism for SNR ages shorter than that of Tycho. 
 
\begin{figure}[!bt] 
  \begin{center} 
    \epsfig{file=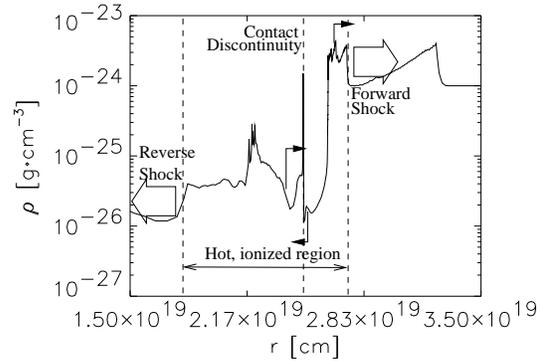, width=7cm} 
  \end{center} 
\caption{
Detail of the structures arising as a result of the interaction of a realistic 
explosion model (PDD has ben chosen for this example) with a CSM modified by a
time-dependent wind (wind model A here). The plot shows the density profile at the
age of SNR 1006, in the region between the ejecta reverse shock
and the wind forward shock. The location of some secondary shocks has been marked with
small arrows.
}   
\label{ebravo-B2_fig8} 
\end{figure}

The interaction of realistic CSM, produced by non-constant presupernova mass loss, 
with the ejecta density profiles obtained from realistic SNIa  
hydrodynamic calculations results in SNR configurations that show a lot  
of structure. Particularly interesting is the presence of several secondary shock 
waves traveling outwards and  
inwards (see Figure \ref{ebravo-B2_fig8}). 
These secondary shocks could have important  
consequences for the overall SNR structure and properties (magnetic  
field, cosmic ray acceleration, hydrodynamic instabilities, ionization
structure, etc.). 
The denser areas in the hot, ionized  
intershock region will display the highest X-ray luminosity (regions  
outlined by the dashed vertical lines). Thermal emission is expected from these  
regions, dominated by lines from solar abundance plasma for the  
shocked CSM (density spike close to the forward shock) or by lines from the elements 
synthesized in the  
SN explosion for the shocked ejecta (density spike between
the contact discontinuity and the reverse shock). 
  
Consistent modelling of the thermal X-ray emission from SNRs remains an  
unsolved problem. An accurate description of nonequilibrium ionization is  
necessary to obtain realistic temperature and luminosity profiles (for a  
discussion, see e.g. \cite{ebravo-B2:bor01}). This is specially difficult for  
the heavy-element ejecta characteristic of SNIa. We are working on the  
coupling of hydrodynamical models with atomic databases to be able to produce spectra  
models that can be used to fit observations and thus take advantage of  
the high sensitivity and spectral resolution of {\sl XMM-Newton} and {\sl Chandra}. 
 
\begin{acknowledgements} 
 
This work has been supported by the MCYT grants EPS98-1348 and AYA2000- 
1785 and by the DGES grant PB98-1183-C03-02. C. B. is very indebted to  
CIRIT for a grant. 
 
\end{acknowledgements}

\end{document}